\documentstyle[12pt]{article}
\newcommand{\m}{\mbox}
\newcommand{\be}{\begin{equation}}
\newcommand{\ee}{\end{equation}}
\newcommand{\bey}{\begin{eqnarray}}
\newcommand{\eey}{\end{eqnarray}}
\title{DIMENSIONAL CROSSOVER FOR THE BOSE -- EINSTEIN CONDENSATION OF AN
IDEAL BOSE GAS\footnote{To be submitted to Physics Letters A}} 
\author{Mario Molina and Jaime R\"ossler\\[10pt] Departamento de
F\'\i sica, Facultad de Ciencias,\\ Universidad de Chile, Casilla
653, Santiago, Chile}
\date{}
\begin{document}
\pagestyle{myheadings}
\maketitle
\vspace{5.0cm}

\newpage

\centerline{\bf ABSTRACT}
\vspace{1.5cm}

\noindent
We study an ideal Bose gas of $N$ atoms contained in a box formed by two
identical planar and parallel surfaces ~$S$~, enclosed by a mantle of height
~$a$~ perpendicular to them. Calling ~$r_0$~ the mean atomic distance, we
assume \mbox{~$ S \gg r_0^2$~,} while ~$a$~ may be comparable to ~$r_0$. In
the bidimensional limit ($ a/r_0 \, \equiv\, \gamma << 1$) we find a {\em
macroscopic} number of atoms in the condensate at temperatures \mbox{~$T\sim
{\cal O}(1/\log(N))$}; therefore, condensation cannot be described in terms of
intensive quantities; in addition, it occurs at temperatures not too low in
comparison to the tridimensional case. When condensation is present we also
find a macroscopic occupation of the low--lying excited states. In addition,
the condensation phenomenon is sensitive to the {\em shape} of ~$S$.  The
former two effects are significant for a nanoscopic system. The tridimensional
limit is slowly attained for increasing $\gamma$, roughly at $\gamma\sim
10^{2}-10^{3}$.

\newpage

\begin{sloppypar}

\section{INTRODUCTION}
%\markboth{left_head}{right_head}

A system of $N$ {\em non-interacting particles} is in a {\em condensed} phase
if a non--negligible fraction of them lie in the ground state; however, for
small values of ~$N$~, say ~$10^3$--$10^5$~, the condensation boundary becomes
blurred. As it is well-known, for an ideal gas obeying Boltzmann statistics,
condensation occurs at vanishingly small temperatures, irrespective of the
dimensionality ~$d$~ of the system
%------------------------------------------------------------------------
(\mbox{~$T \,\sim\, {\cal O}\{1 /N^2\}$~} for \mbox{$d\,=\, 1$}; \mbox{~$T
\,\sim\, {\cal O}\{1/N\} $~} for \mbox{$d\,=\,2$} ~and~ \mbox{$T\,\sim\, {\cal
O}\{1/N^{2/3}\} $~} for \mbox{$d \,=\, 3$}). In contrast, for Bose statistics,
condensation begins at a finite temperature in three-dimensional systems,
while it is absent in one- and two-dimensional systems~\cite{books}, at least
in the {\it mathematial limit} \mbox{~$N \to \infty$} (\mbox{~$T \,\sim\,{\cal
O}\{1/N\}$~} for \mbox{$d\,=\, 1$} ~and~ \mbox{~$T \,\sim\, {\cal O}\{1
/\log(N)\}$~} for \mbox{$d\,=\,2$}, see Ref.~\cite{hohenberg}). 
%------------------------------------------------------------------------
The presence of an strong enough external potential can, however, bring about
a Bose-Einstein condensation (BEC) in two and even one dimensions, at a finite
temperature\cite{bagnato}.

Since Bose and Einstein predicted the condensation phenomenon, there has been
a great interest into achieving such a state in an actual system. The
existence of superfluidity in \mbox{~$^{4}He$~} has been considered a sort of
corroboration of BEC. However, superfluid Helium is a poor example of a
non--interacting particle system, since the mean atomic separation is
comparable with the range of the atomic potential. Therefore, a complete
explanation of superfluidity must rely in Many-Body Theory. 

The goal of obtaining BEC in a gas of nearly non-interacting atoms was
recently achieved by Anderson {\em et al}.\cite{anderson}\ who combined the
techniques of laser and evaporative cooling of magnetically trapped atoms,
implementing the latter one by means of an ingenious ``time orbiting
potential'' magnetic trap. They obtained temperatures of \mbox{~$\sim 10^{-7}\
K $~} for a sample of a few thousand atoms trapped in a region of linear
dimensions \mbox{~$\sim 10^{-5}\ m$~} and were thus able to achieve BEC. The
nanoscopic character of the experiment motivates a more detailed study of BEC
for such small systems.
\vspace{0.4cm}
 
The purpose of this work is to analyze the condensation process of an ideal
Bose gas as a function of the dimensionality of the system and the shape of
the container.  Since a rigorous infinite two--dimensional gas is not a
realistic system, we shall consider here a {\it true tridimensional gas}
enclosed in a finite container; our aim is to understand how the properties of
the gas change as the container is shrunken in one of its dimensions (say,
$z$), expecting a continuous passage from tri to bidimensional behavior.
%------------------------------------------------------------------------
In fact, at enough small temperatures, when the de Broglie wavelenght
\mbox{$\lambda(T)$} becomes smaller than the container width, there is not
enough thermal energy to excite the degree of freedom associated to the ~$z$~
direction. Therefore, one dimension ``freezes'' and our system behaves like an
``effective'' two--dimensional gas, in spite of its actual tridimensional
nature. If the container width is comparable to the interatomic mean distance,
condensation must occur after the bidimensional regime has been established.
%------------------------------------------------------------------------
\vskip 0.4cm

Interest in this problem dates back to London's suggestion 
to use BEC as a qualitative model for understanding superfluidity
in He$^{4}\ $\cite{london}. Experiments showed that He$^{4}$ remained a
superfluid when forced to flow inside narrow capillaries. On the other hand,
superconductivity was shown to persist in thin films. This prompted the study
of ``superproperties'' in finite geometries. In one of the first
studies\cite{mills}, an ideal bose gas was enclosed in a box of finite volume
($L\times D\times D$) and the number of particles in the ground state was
computed numerically as a function of the temperature. Large deviations from
the bulk result were found in the one-dimensional limit ($L>>D$).  Attempts to
ascertain the dependence of the superfluid transition temperature for He$^{4}$
on the dimensions of narrow channels\cite{khorana}, were followed by
discussions on the ambiguity of the concept of a transition temperature in a
finite geometry, where a rigorous phase transition is
absent~\cite{trainor,krueger2}. Vanishing of quasiaverages in partially finite
geometries (one or more dimensions finite while one or more dimensions extend
to infinity) led to the suggestion\cite{krueger} that these geometries were
not good approximations to thin films and pores found in laboratories.  For a
strictly finite geometry, BEC was shown to occur for an ideal
gas\cite{krueger}.   

While in most of these earlier works, the goal was to achieve a qualitative
understanding of superfluid properties, we are now genuinely concerned with
describing the behavior of a finite number of nearly-independent bosons,
inside a confining potential, such as in the recent experiments mentioned at
the beginning\cite{anderson}. The recent advances in nanotechnology and laser
optics allow researchers now to manipulate molecules and their environment at
the mesoscopic and nanoscopic levels, rendering the investigation of quantum
effects in a system with a finite number of components (i.e., away from the
thermodynamic limit), an interesting field of study. 

In the present work we shall address the problem of crossover between two and
tridimensional behaviour in BEC. We shall give both, exact numerical and
asymptotic analytical results. For the sake of completeness we shall refresh
some old results. However, our anlytical expressions will give a compact and
clear sight of the phenomenology under consideration; in particular, they show
the non--intensive character of BEC near the two--dimensional limit, giving
simple estimations for the temperature of condensation (we remember that the 
latter temperature is not sharply defined in this case). We shall also clarify
the {\it effect of geometry} in BEC, which can be summarized by  some
characteristic constants associated with the shape of the container, listing
them for some particular cases; incidentally, the different boundary
conditions will be considered like peculiarities of the topology of the
container, affecting the values of former constants. The role of lower excited
states and its macroscopic occupance shall be also addressed. Finally, we
shall also analyze the continuous transition between two and tridimensional
geometries. 

\section{THE MODEL}

We study an ideal Bose gas of $N$ atoms contained in a box formed by two
identical planar and parallel surfaces ~$S$~ (say, in the ~$x,y$~ plane),
enclosed by a mantle of height ~$a$~ perpendicular to them. We call ~$r_0$~
the mean atomic distance, and introduce the parameter \mbox{~$\gamma=
a/r_0$~.} Since the container volume is \mbox{~$S a$~,} it holds \mbox{~$r_0^3
\,=\,S a/N$~.} We assume that \mbox{~$S \gg a^2 $~} (or equivalently \mbox{~$N
\gg \gamma^3 $~),} while ~$a$~ may be comparable to ~$r_0$. 

We introduce a characteristic temperature for the ideal gas, associated to the
energy of a particle whose de Broglie wavelenght $\lambda(T)$ is comparable to
the mean intermolecular distance ~$r_0$, 

\be
T_0 \;\equiv\; \frac{h^2}{2 \pi m k_B r_0^2} 
\label{T0}
\ee

\noindent
where $m$ is the mass of a particle and ~$h$~, ~$k_B$~ are the Planck and
Boltzmann constants, respectively. For \mbox{~$T < T_0$~} the bidimensional
behaviour can be associated with \mbox{~$\gamma \sim {\cal O}(1)$~,} while
\mbox{~$\gamma \gg {\cal O}(1)$~} corresponds to the tridimensional limit. We
also define the ``reduced'' temperature \mbox{~$\tau \,=\, T/T_0$~;} now the
(tridimensional) BEC critical temperature is
\mbox{~$\tau_c\,=\,0.527201068760\dots$~;} see Ref.~\cite{books}. 

The energy dispersion relation for a particle enclosed there can be expressed
like 

\be
E(\nu,\ell) \,=\, k_B T \left\{\eta \xi_{\nu}^2 \,+\, \sigma \ell^2 \right\}
\label{energia} 
\ee 

\noindent
here ~$\sigma \,\equiv\, \pi/[\tau \gamma^2 (q+1)^2]$~, where ~$q=0$ for
ciclic boundary conditions (CBC) and ~$q=1$~ for impenetrable boundary
conditions (IBC) in the ~$z$~ direction respectively;
\mbox{~$\ell\,=\,0,\,\pm 1,\,\pm 2,\dots$} for CBC, while
\mbox{~$\ell\,=\,1,\,2,\,3\dots$} for IBC. \mbox{~$\eta\,\equiv\, \pi/\tau
N_s$~,} where \mbox{~$N_s\,=\,N/\gamma\,=\, S/r_0^2$~} is the number of
particles in a neighbourhood ~$r_0$~ of container base ~$S$.  The numbers
\mbox{~$\{\xi_\nu\}$~} describe the spectrum of the bidimensional Helmholtz
equation for the surface ~$S$; they {\it depend only on the shape of} ~$S$ and
the boundary conditions used for the \mbox{~$x,y$~} plane. We choose the
ground state as energy zero; therefore we shift \mbox{~$\xi_{\nu}^2 \to
\xi_{\nu}^2 -\xi_{0}^2$~,} while \mbox{~$\ell^2\to \ell^2-1$~} for IBC in the
~$z$~ direction. 

We shall use the asymptotic condition for the bidimensional energy

\be
D(\xi)\ d\xi \;\equiv\; \left\{\mbox{Number\ of}^{}\ \xi_{\nu} \in [\xi\,, \xi
+ d\xi]\right\} \;=\; \left(2\pi \xi - C_s\right)\, d\xi 
\label{D(E)}
\ee

\noindent
valid for ~$\xi\; d\xi \gg 1$~. Here ~$C_s$~ is a constant that depends of the
shape of ~$S$~; therefore, the contribution \m{~$-C_s$~} in Eq.~(\ref{D(E)})
represents the finite size corrections to the density of
states~\cite{molina}; for example, \m{~$C_s\,=\,0$} for a square with
CBC, \m{~$C_s\,=\,2$} for a square with IBC, and \m{~$C_s\,=\,\sqrt{\pi}$}
for
a circle with IBC. The mean occupation of the state \m{~$E(\nu,\ell)$~} is
given by

\be
{\cal N}(\nu,\ell) \,\equiv\, N \cdot n(\nu,\ell) \,=\, \left\{ \zeta
\exp\left[\eta \xi_{\nu}^2 \,+\, \sigma \left(\ell^2-q\right) \right] \,-\,1
\right\}^{-1}~. 
\label{N_ocup}
\ee

\noindent
here \mbox{~$\zeta \, \equiv\, \exp(-\mu/k_B T)$~,} where ~$\mu$~ is the
chemical potential. According to the former relation, the ground state mean
occupation is \mbox{~${\cal N}_0\,\equiv\,N\cdot n_0 =1/(\zeta-1)$~}. 

For IBC in the ~$z$~ direction we shift the index \mbox{~$\ell\to
\ell-1$~,} in order to associate ~$\ell=0$~ to the ground state. Due to the
fact that \mbox{~$N \gg \gamma^3$~,} it holds that
\mbox{$n(\nu,\ell \neq 0)\,\sim\,{\cal O}\{1/N\} $,} while, for
\mbox{~$\ell=0$~} the lower
excitations in the \m{~$\{x,y\}$~} plane may have a macroscopic population.
In fact, when a macroscopic amount of particles lie in the ground state,
relation~(\ref{N_ocup}) leads to

\be
 n(\xi_{\nu},0)\;\approx\; \left\{\frac{1}{n_0}\,+\,
\frac{\pi\gamma}{\tau}\left(\xi_{\nu}\right)^2\,+\,{\cal O}(\frac{1}{N})
\right\}^{-1} 
\label{N_aprox}
\ee

The equation for ~$\zeta$~ is \m{~$1 = \sum_{\nu, \ell} \, n(\nu, \ell)$.~}
In
order to study the condensation phenomenon, it is necessary to separate the
former
sum into contributions from states with macroscopic
(\mbox{~$ n(\nu, \ell =
0)\,\sim\, {\cal O}(1)$~}) and microscopic (\mbox{~$ n(\nu, \ell\neq
0)\,\sim\, {\cal O}(1/N)$~)} occupancy. Only in the latter contributions we
can safely approximate \mbox{~$\sum_{\nu}\, \to\, \int d\xi D(\xi)$~} since
then \m{~$n(\nu,\ell)$~} is a smooth function of ~$\nu$. But the macroscopic
contributions \m{~$n(\nu,0)$~} must be summed explicitly, at least for the
lower energy states. Over a high enough value of ~$\nu$~ it is safe
to replace this sum by an integral; defining \m{~$\xi_L \,<\,R\,<\, \xi_{L+1}
$~,} \m{~$R \gg 1$~,} the equation for ~$\zeta$~ becomes  

\bey
 1 & = & \frac{1}{N\,(\zeta\,-\,1)} \;+\; \sum_{\nu=1}^{L} n(\xi_{\nu},0) 
\nonumber\\ 
   &   & \;+\;\frac{2}{(q+1) N} \sum_{\ell=1}^{\infty}
\int_0^{\infty}  d\xi\, D(\xi) \left\{\displaystyle \zeta \exp\left[\eta \xi^2
\,+\,\sigma \ell(\ell+2q) \right] \,-\,1\right\}^{-1} \;+\; 
\nonumber\\ 
   &   & 
\frac{1}{N} \int_{R}^{\infty} d\xi\, D(\xi) \left[\displaystyle
\zeta\exp\left(\eta \xi^2 \right) \,-\,1\right]^{-1} 
\label{cond_z} 
\eey

\noindent
where \m{~$1/[N\,(\zeta\,-\,1)] \;\equiv\; n_0 $~} is the fractional ground
state occupation. The bulk contribution to \m{~$D(\xi)$~} can be integrated in
a closed form; the finite size correction to \m{~$D(\xi)$~} gives a negligible
contribution, except for the case when condensation has begun and
\m{~$\ell=0$~;} in the latter case we use Eq.~(\ref{N_aprox}). Thus, the
equation for ~$\zeta$~ takes the form

\bey
 1 & = & \frac{1}{N(\zeta\,-\,1)} \;+\; \frac{1}{N}\, \sum_{\nu=1}^L
\,\left[\zeta\exp\left(\eta \xi_{\nu}^2\right)\,-\,1\right]^{-1} \;-\; 
\nonumber\\ 
& & 
\frac{\tau}{\gamma}\,\log\left[1\;-\;\frac{1}{\zeta}\exp\left(-\eta\,R^2
\right)\right] \;-\;
\nonumber\\ 
& & \frac{2 \,\tau}{(q+1)\, \gamma} \sum_{\ell=1}^{\infty}
\log\left\{1\;-\;\frac{1}{\zeta}\exp\left[-\sigma\ell(\ell+2q)\right]\right\}
 \;-\nonumber\\
& & C_s \sqrt{\frac{\displaystyle \tau n_0}{\pi \gamma}}
\arctan\left[\frac{1}{R}\sqrt{\frac{\tau}{\displaystyle \pi\gamma n_0}}\right]
\label{eq_z}
\eey

This expression is suitable for numerical computation. In practice, we have
summed over the \m{~$L=9000$~} lower states in the case of a rectangular base,
and \m{~$L=1241$~} states for the circle; for such values of ~$R$~ the error
on approximating the sum by an integral is negligible. In addition, for a
given ~$L$~, the specific choice of ~$R$~ in the range \m{~$\xi_L \,< \, R
\,<\, \xi_{L+1}$~} is also irrelevant (the associated variation in ~$n_0$~ is
lower than \m{~$0.4\times 10^{-4}$}). 
\vskip 0.5cm

\large{\bf Some Approximate Results}
\normalsize
\vskip 0.2cm

Before showing and discussing the numerical results, it is useful to obtain
some asymptotic relations for the case \m{~$n_0\, \gg\; 1/N$~} and a not too
large value for ~$\gamma$~, say, \m{~$\sigma \,=\,\pi/[\tau\gamma^2(1+q)^2]\gg
1/[N n_0]$}. Then, it follows from Eq.~(\ref{N_ocup}), \m{~$\zeta\,=\, 1\,+\,
1/(N n_0) \,\approx\, 1$~.} In addition, replacing Eq.~(\ref{N_aprox}) into
Eq.~(\ref{eq_z}) and taking the limit \mbox{~$R \,\to\, \infty$~} we obtain

\bey
 1 \;=\;  n_0 & \;+\;& \sum_{\nu =1}^{L} \,\frac{\tau\; n_0}{\left(
\displaystyle \tau \,+\; \pi \gamma\, n_0\, \xi_{\nu}^2 \right)} \;-\;
\frac{\tau}{\gamma}\,\left[ \log\left(\frac{\pi\gamma\, R^2}{\tau\, N} \right)
\;-\; \Phi(\sigma) \right]
\nonumber\\ 
\mbox{where} & & \hspace{0.2cm} \Phi(\sigma)\;=\;\frac{-\,2}{(q+1)}\,
\sum_{\ell=1}^{\infty}
\log\left\{1\;-\;\exp\left[-\sigma\ell(\ell+2q)\right] \right\} 
\label{eq_z2}
\eey
\vskip 0.5cm

{\it (i)}~ {\bf Small Temperature Case--}~~ We shall assume that condensation
is well established, and that the inequality \m{~$n_0 \gg \tau/\gamma$~}
holds. We expand in a Taylor series the second term of Eq.~(\ref{eq_z2}),
concluding the following relation for ~$n_0$~
 
\bey
0 &\approx & n_0^2 \,-\,n_0 \,+\,
\frac{n_0\,\tau}{\gamma}\left[\Phi(\sigma)\,+\, 
\log\left(N_s \tau\alpha \right)\,\right] \,-\,\left(
\frac{\tau}{2 \gamma}\right)^2 \kappa \nonumber\\
\mbox{where} & & \hspace{2cm} \kappa\;\equiv\; \frac{4}{\displaystyle \pi^2}
\left[\sum_{0<\xi_{\nu}<R}\left(\frac{1}{\xi_{\nu}}\right)^4
\,+\,\frac{\pi}{\displaystyle R^2} \,-\,\frac{C_s}{\displaystyle 3 R^3}
-\right]_{R\to \infty} \nonumber\\
\mbox{and} & & \alpha\;\equiv\; \frac{1}{\pi}\exp\left[\frac{1}{\pi}\,
\sum_{0<\xi_{\nu}<R}\left(\frac{1}{\xi_{\nu}}\right)^2 \,-\, 2 \log(R)\,-\,
\frac{C_s}{\pi R} \right]_{R \to \infty} 
\label{n0ap1}
\eey

Solving for ~$n_0$~ we obtain

\bey
n_0(\tau) & \approx & \frac{1}{2} \left\{\, m \,+\, \left[ m^2\;+\;
\kappa \left(\frac{\tau}{\gamma} \right)^2 \, \right]^{1/2}\right\}\nonumber\\
\mbox{where} & & \hspace{1cm} m(\tau)\;\equiv\; 1\,-\,\frac{\tau}{\gamma}
\left[\, \log(\alpha \tau N_s) \,+\, \Phi(\sigma)\, \right]
\label{n0ap2}
\eey

An estimation of the accuracy of this approximation is\\ \mbox{~$\Delta
n_{0}=n_{0} - \{n_{0}\}_{exact}\,\sim\, C \tau^3/(\gamma^3\, n_{0}^{2})$~,}
where \mbox{~$C\,\sim\, 0.2$~,} at least for square and circle geometries.
Thus, the error strongly decreases as $n_{0}$, $N_{s}$ or $\gamma$ increases.
However, the error is very small for \m{$n_{0}\,= \, 0.2$}, where we have
\m{~$\Delta n_{0}\;\leq\; 0.01$~} if \m{~$N \geq 10^{4}$,} \m{~$\gamma\geq
1$.} When \m{~$N>10^8$~} the error \m{~$\Delta n_{0}\;\leq\; 0.003$~,} even
for a very small amount of condensate, \m{~$n_0 \,\geq\, 0.08$~.} The accuracy
of Eq.~(\ref{n0ap2}) decreases if the shape of ~$S$~ is very ``oblong''. It
is important to remark that the constants ~$\alpha$~ and ~$\kappa$~ only
depend on {\it the shape of surface} ~$S$~; we include a table with the values
of these constants for a circle and a rectangle \mbox{~$b_1 \times\, b_2$~;}
in
the latter case we use different boundary conditions and values for the
ratio
\m{~$\rho\,=\,b_1/b_2$~} (e.g. \m{~$\rho\,=\,1$~} corresponds to a square).

As a consequence of these geometric constants, we conclude that condensation
is inhibited when the shape of ~$S$~ is very oblong; the latter one is
consistent with the absence of condensation in a nearly one--dimensional
container~\cite{books}, in accordance with our exact calculations (see
Fig.~1). 
\vskip 0.5cm

{\it (ii)}~ {\bf Intermediate Temperature Case--}~~ Using Eq.~(\ref{eq_z2}) in
the case \m{~$1/N \,\ll\, n_0\, \ll\, 1$~,} when condensation is scarcely
incipient, we have concluded the asymptotic relation

\be
n_o \sim\, {\cal O}\left\{\gamma \,\tau_I \left(\frac{1}{N}\right)^{(\tau-
\tau_I)/\tau} \; \exp\left[-2\gamma(1+q)\sqrt{\frac{\tau}{\pi}~}\right]
 \right\}
\label{dudoso}
\ee  

\noindent
where ~$\tau_I\,=\, \gamma/\log(N)$~ roughly corresponds to the temperature
threshold for condensation appearance. In this way, the typical temperature
for condensation in a quasi--bidimensional system roughly corresponds to \m{~$
\tau_I \sim \tau_c/ \log(N)$~,} where ~$\tau_c$~ is the tridimensional
critical temperature for BEC; the latter one implies that, typicaly, the
quasi--bidimensional condensation occurs at temperatures one order of
magnitude below the tridimensional case. 
\vskip 0.5cm

\large{\bf Results and Discussion}
\normalsize
\vskip 0.2cm

We have solved Eq.~(\ref{eq_z}), determining \m{~$n_0(\tau)$~} for several
values of ~$N$~, ~$\gamma$~ and different shapes of ~$S$: a circle, rectangles
with different flatness ~$\rho$~. We have always used IBC in the normal
direction with respect to ~$S$. We display our numerical results in
figs.~{
1--4}. In what follows we summarize the principal consequences of these
results.
\vskip 0.6cm

{\it (i)}~ The shape of the base ~$S$~ influences the condensation phenomenon,
speciallly in the nearly bidimentional limit, \m{~$\gamma \,\sim\, {\cal O}\;
\{ 1\}$~.} This effect is also enhanced for relatively small values of ~$N$~.
In Fig.~1 we illustrate this effect for \m{~$\gamma\,=\,1$~} and
\m{~$N\,=\,10000$~.} We see that a IBC--square base yields the larger values
of \m{~$n_0(\tau)$~} (at least for not too small values of ~$n_0$); however,
the circle gives only slightly lower values for ~$n_0$ in comparison to the
IBC--square, while the CBC--square curve \m{~$n_0(\tau)$~} lies below the two
former ones. Finally the lower values of \m{~$n_0(\tau)$~} in Fig.~1
correspond to IBC--rectangles; these curves descend on increasing the flatness
~$\rho$~, as the system approaches to a one--dimensional behaviour. The latter
one is in accordance with the absence of BEC in that limit~\cite{books}.

This ``shape effect'' on \m{~$n_0(\tau)$} is important in the 2--d limit,
diminishing upon increasing $\gamma$ or $N$.  Therefore, it can only be
observed in truly nanoscopic systems. In the following item we shall explain
this behaviour.
\vskip 0.4cm

{\it (ii)}~ According to Eq.~(\ref{N_aprox}), the low--lying excited states
also have a macroscopic occupancy once condensation begins. In Fig.~2 we
display the occupancy of the lower seven excited levels in the case of
circular base, \m{~$N=10000$~} and \m{~$\gamma =1$~.} The associated quantum
numbers are \m{~$\nu\,=\,(m,j)=$~} (0,1),~ (1,1),~ (2,1),~ (3,1),~ (1,2),~
(0,2)~ and (2,2). Here ~$m$~ is associated to the axial component of angular
momentum ~$L_z$~, while ~$j$~ enumerates the eigenenergies for a given ~$m$. 
Obviously, while ~$m=0$~ is not degenerate, \m{ ~$m\neq 0$~} has degenerancy
two. The effect of degeneracy becomes evident in the asymptotic behaviour
\m{~$\tau\rightarrow \infty$~,} where the occupancy of non--degenerate states
is one--half of the degenerate ones. The maximal occupancy of an excited level
corresponds to (1,1) (say, the first excited state), giving
\m{~$n_{(1,1)}(\tau_{Max}=0.13)\;=\; 0.093$~.} 

In the general case, the macroscopic occupancy of excited levels is enhanced
as ~$\gamma$~ or ~$N$~ degreases. An approximate relation for the maximal
occupation of these excited levels is given by

\be
n_{\nu}(\tau_{Max}) \;\approx \; \frac{D_{\nu}}{\displaystyle \theta
\xi_{\nu}^2 } \hspace{1.2cm} \m{with}  \hspace{1.2cm} \theta\,=\,
\pi\, \log\left[\frac{N \alpha}{\log\left(N_s \alpha \right)}\right]
\label{n_ex_M}
\ee
 
\noindent 
and \m{~$\tau_{Max}\,\approx\, \pi \gamma/\theta$~;} here ~$D_{\nu}$~ is the
degeneracy. Thus, this maximal occupancy depends on ~$N$~ like \m{~$1/
\log(N)$~} (thus increasing as ~$N$~ decreases, being appreciable in truly
nanoscopic systems), and it is inversely proportional to the bidimensional
excitation energy \m{~$E(\nu,0) -E(0,0)$~} (see Eq.~(\ref{energia})~). We
remark that Eq.~(\ref{n_ex_M}) is valid under the condition \m{~$\gamma \sim
1$~;} for larger values of ~$\gamma$~ the occupancy of the excited levels
strongly decreases (for example, using the same ~$N_s$~ of Fig.~2, but
\m{~$\gamma\,=\,10$~,} we obtain a maximal occupancy
\m{~$n_{(1,1)}(\tau_{Max}=0.47)\;=\; 0.037$~}).

According to relation~(\ref{n_ex_M}), the {\it macroscopic} occupancy of the
lower excited levels is strongly affected by the particular sequence
\m{~$\{\xi_{\nu}\}$~,} which, in turns, depends of {\it the shape of} ~$S$~.
The latter one implies that the occupancy of the ground state is also affected
by the base shape. In particular, the amount of condensate ~$n_0$~ increases
if the gap between the ground state and the first excited levels increases,
i.e., when the level density decreases. For a fixed base area but different
shapes, we can modify the former magnitudes and therefore modify ~$n_0$.

%%%>>>>>>REVISAR NUMARICAMENTE LO QUE SIGUE:
 
In order to obtain a further feeling for the effect of the shape in the amount
of condensate \mbox{~$n_0(\tau)$~,} it is very illustrative to note the
following numerical result: The sum of occupancies of the ground state and the
first excited level is roughly the same for the circle, the CBC--square and
the IBC--square; the latter one, near the theshold temperature of
condensation. Therefore, the shape effect on ~$n_0$~ roughly corresponds to
the difference in the populations of the first excited levels. Although this
estimation is very crude, and certainly not applicable to arbitrary shapes
(specially if one of them yield a quasi--degenerate first excited level), it
gives a first estimate for the shape effect on the thermodynamics of the
condensate. 

%% >> REESCRIBIR ESTE PARRAFO USANDO LA FORMA DE ALPHA 

The effect of shape in condensation is also evident from
relations~(\ref{n0ap1}) and~(\ref{n0ap2}), since the geometrical constant
~$\alpha$~ increases as the density of levels in the neighbourhood of ground
state decreases (see Eq.~~(\ref{n0ap2})). An increase in ~$\alpha$~ implies a
decrease of \m{~$n_o(\tau)$~,} in accordance with Eq.~(\ref{n0ap2}). 

The results of Fig.~1 is consistent with this explanation. In addition, since
the occupancy of the low lying excited levels is only important for small
values of $\gamma$ or $N$, the effect of shape in the ground state occupanncy
\m{~$n_0(\tau)$~} is appreciable under the same conditions. Therefore, this
``shape effect'' can only be observed in truly nanoscopic systems. 

The latter observation is consistent with the relation between the density of
states \m{~$D(E)$~} an the system dimensionality. For example, in a fully
tridimensional system, \m{~$D(E)\,\sim\, \sqrt{E}$~,} vanishing as
\m{~$E\,\to\, 0$~;} this low density is responsible of a genuine BEC at a
relatively hight temperature. In the bidimensional case, \m{~$D(E)\,\sim\, \{
\m{\it Constant} \}$~;} while the onedimensional case corresponds to
\m{~$D(E)\,\sim\, 1/\sqrt{E}$~,} diverging as \m{~$E\,\to\, 0$~.} Accordingly,
the intermediate density of states of the bidimensional case is consistent
with condensation phenomenon at temperatures one order of magnitude below the
tridimensional case. Finally, in the one--dimensional case, the high density
of states near the ground state is in accordance with the absence of
condensation in that case. 
\vskip 0.4cm

{\it (iii)}~  According to relation~(\ref{n0ap2}), the condensation phenomenon
is highly non--intensive in the case of a small ~$\gamma$~, due to the
dependence of \m{~$n_0(\tau)$~} on \m{~$\log(N)$~.} Moreover, using that
relation we can estimate an ``effective critical temperature''
\m{~$\tau_{eff}\; \approx\; \gamma/\log[\alpha N /\log(N)]\;\sim\;
\gamma/\log(N)$~;} this estimation is consistent with ~$\tau_I$~ of
Eq.~(\ref{dudoso}) and ~$\tau_{Max}$~ of Eq.~(\ref{n_ex_M}).

In figs.~3 and 4 we plot \m{~$n_0(\tau)$~} for ~$N=10^4$  and$~10^8$ for
the case of a circle; in each figure we take different values of ~$\gamma$.
In general terms, these figures confirm that condensation is inhibited on
decreasing ~$\gamma$~ or increasing ~$N$. Inspection of these figures
confirm the {\it non--intensive character} of \m{~$n_0(\tau)$~.} For example,
the curves \m{~$\gamma\,=\,1$~} of Fig.~3 and \m{~$\gamma\,=\,2$~} of Fig.~4
nearly coincide, in agreement with the fact that \m{~$\tau_{eff}\;\sim\;
\gamma/\log(N)$~} has the same value in both cases. 

The condensation phenomenon has a nearly 2--d character for small
values of ~$\gamma$~, as long as the excitations in the (``thin'') ~$z$~
direction are very improbable at temperatures compatible with condensation of
a macroscopic number of particles.  An upper bound for this 2--d character can
be roughly estimated as \mbox{~$\gamma \leq 2$~} for a nanoscopic system
\mbox{($N_s \sim 10^4 \leftrightarrow 10^8$)}. The tridimensional limit is 
slowly attained as ~$\gamma$~ increases; roughly, for \m{~$\gamma \sim 10$~}
if \m{~$N=10^4$~,} and \m{~$\gamma \sim 50$~} if \m{~$N=10^8$~.} In this way,
on increasing ~$\gamma$~, the tridimensional limit is attained faster for a
small number of particles; this result seems natural, since for a fixed value
of ~$\gamma$~ the base width \m{($\sim \,\sqrt{S}$)} and container height
($a$) may be comparable if ~$N$~ is not too large, but  \m{($\sqrt{S} \,\gg\,
a $)} for a greater value of ~$N$. 

Another important results illustrated in Fig.~3 are: {\it (a)}~~the finite
size effects, which yield a larger amount of condensate for ~$\gamma=20$~ in
comparison to the tridimensional case. {\it (b)}~~The condensation phenomenon
begins gradually for low values of ~$N$~ and ~$\gamma$~, in close accordance
with the qualitative result~(\ref{dudoso}); while larger values of  ~$N$~ and
~$\gamma$~ yield a more sharp phase transition.

%%%%\large{\bf Discussion}
%%%%\normalsize

\vspace{2.0cm}

\centerline{\bf ACKNOWLEDGEMENTS}
\vspace{0.8cm}

This work has been partially supported by FONDECYT project 1950655
(1995--1997).

\newpage
{\centerline{\bf Figure Captions}}
\vspace{0.5cm}

\noindent
{\bf Figure 1:}\ \ Number of particles in the ground state as a function
of the reduced temperature for several different shapes of the container
base. In all cases, we have IBC, except for the lower square, where CBC are
used. ($\gamma = 1, N = 10^4$).
\vspace{0.5cm}

\noindent
{\bf Figure 2:}\ \ Occupancy of the lower seven excited states
for the case of a circular base. Each state is labelled by $(m, j)$ where
$m$ is associated to $L_{z}$ while $j$ enumerates the eigenenergies
corresponding to a given $m$. ($\gamma = 1, N = 10^4$).
\vspace{0.5cm}

\noindent
{\bf Figure 3:}\ \ Circular base:\ Number of particles in the ground state
as a function of the reduced temperature, for different values of $\gamma$.
($N = 10^4$).
\vspace{0.5cm}

\noindent
{\bf Figure 4:}\ \ Same as in fig.3, but for $N=10^8$.
\newpage
%.....................................................................
\begin{center}
\begin{tabular}{|l|l|l|l|l|}
\hline
Surface & $\rho$       & Boundary   & $\alpha$ & $\kappa$\\
        &  (Rectangle) & Conditions &          &         \\
\hline \hline  
Circle    & --  &  I.B.C. & 0.4019...    & 2.4967... \\ \hline
Square    & 1  &  I.B.C. &  0.35106...   & 2.280612... \\ \hline
Rectangle & 1.5 &  I.B.C. & 0.391197...  & 2.854037... \\ \hline
Rectangle & 2   &  I.B.C. & 0.489620...  & 4.159725... \\ \hline
Rectangle & 3  &   I.B.C. & 0.888309...  & 8.319915... \\ \hline
Rectangle & 5   &  I.B.C. & 3.76785...   & 22.1509... \\ \hline
Rectangle & 10 &   I.B.C. & 232.3111...  & 87.65845... \\ \hline
Square    & 1   &  C.B.C. & 0.7247...    & 2.442575... \\ \hline
Rectangle & 1.5 &  C.B.C. & 0.80982...   & 2.995687... \\ \hline
Rectangle & 2   &  C.B.C. & 1.02498...   & 4.274506... \\ \hline
Rectangle & 5 &  C.B.C.   & 9.48733...   & 22.23856... \\ \hline
Rectangle & 10 &  C.B.C. & 891.3993...   & 87.8829... \\ \hline
\end{tabular}
\end{center}
\newpage
%%%%%%%%%%%%%%%%%%%% fig 1  %%%%%%%%%%%%%%%%%%%%%%%%
\begin{figure}[p]
\begin{center}
\leavevmode
\hbox{	
\includegraphics{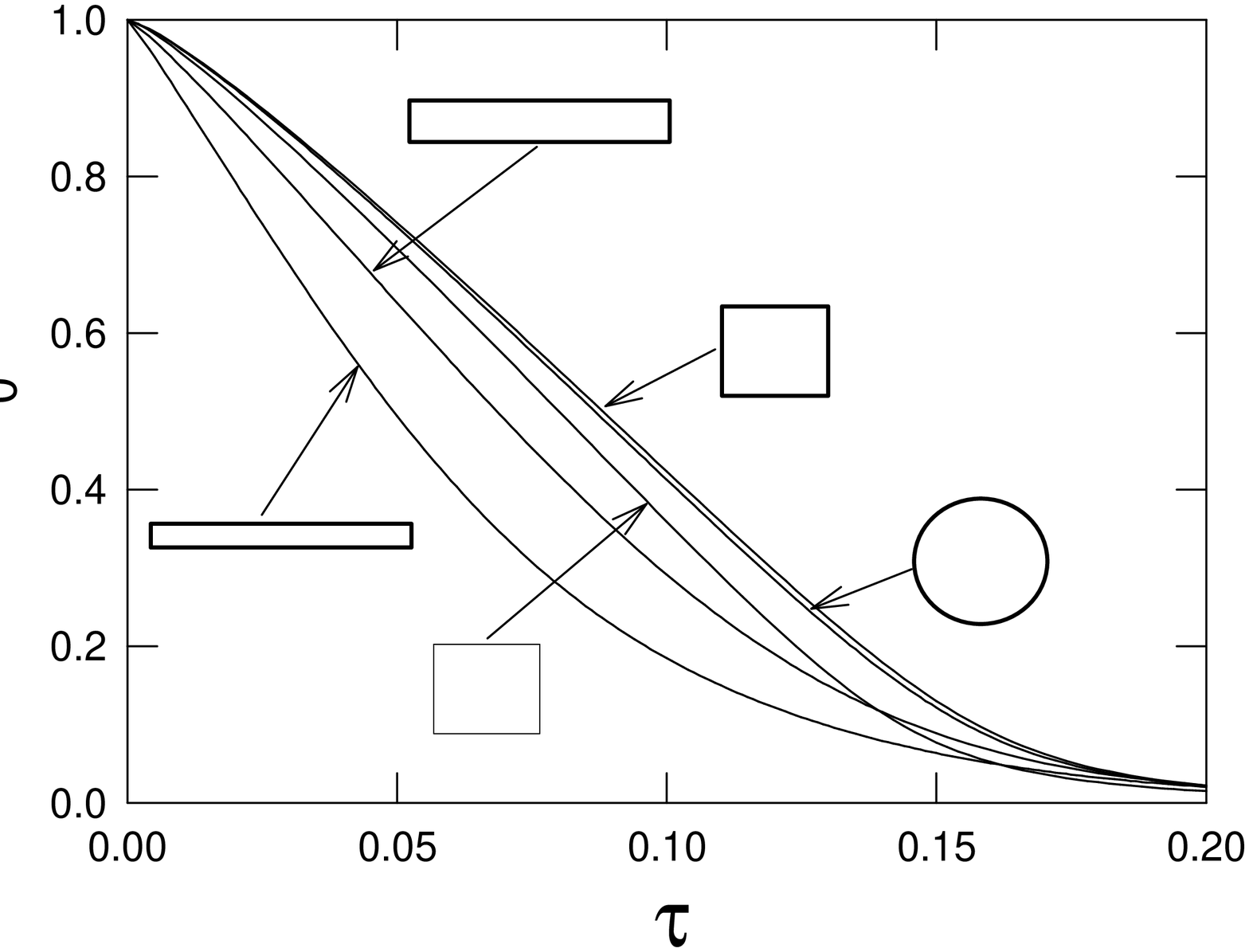} }    	
\end{center}
\vspace{2.3 in}
\label{figure1}
\vspace{6.0cm}
\centerline{}
\end{figure}
%%%%%%%%%%%%%  fig 2 %%%%%%%%%%%%%%%%%%%%%%%%%%%%%%%
\begin{figure}[p]
\begin{center}
\leavevmode
\hbox{	
\includegraphics{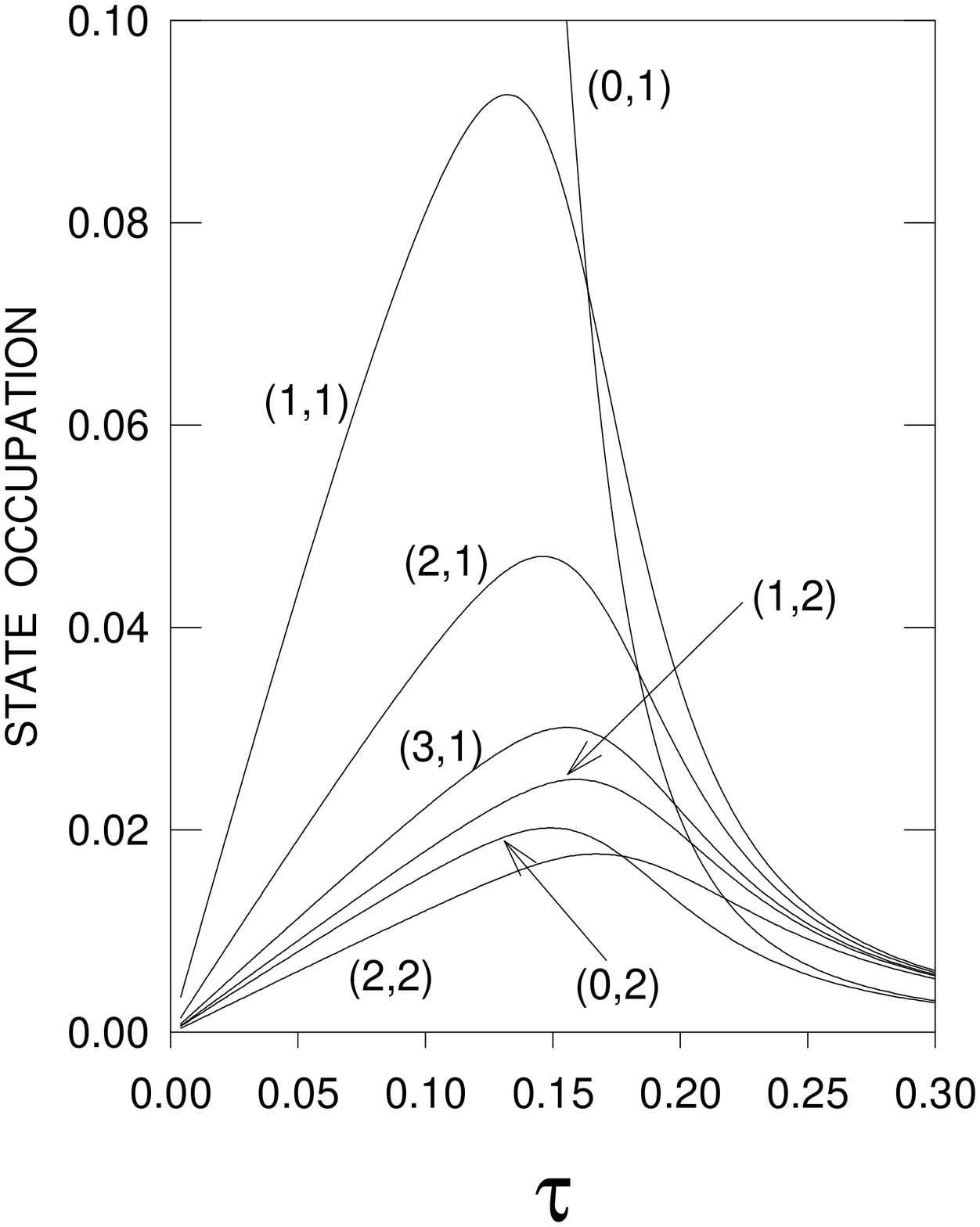} }    	
\end{center}
\vspace{2.3 in}
\label{figure2}
\vspace{6.0cm}
\centerline{}
\end{figure}
%%%%%%%%%%%%  fig 3 %%%%%%%%%%%%%%%%%%%%%%%%%%%%%%
\begin{figure}[p]
\begin{center}
\leavevmode
\hbox{	
\includegraphics{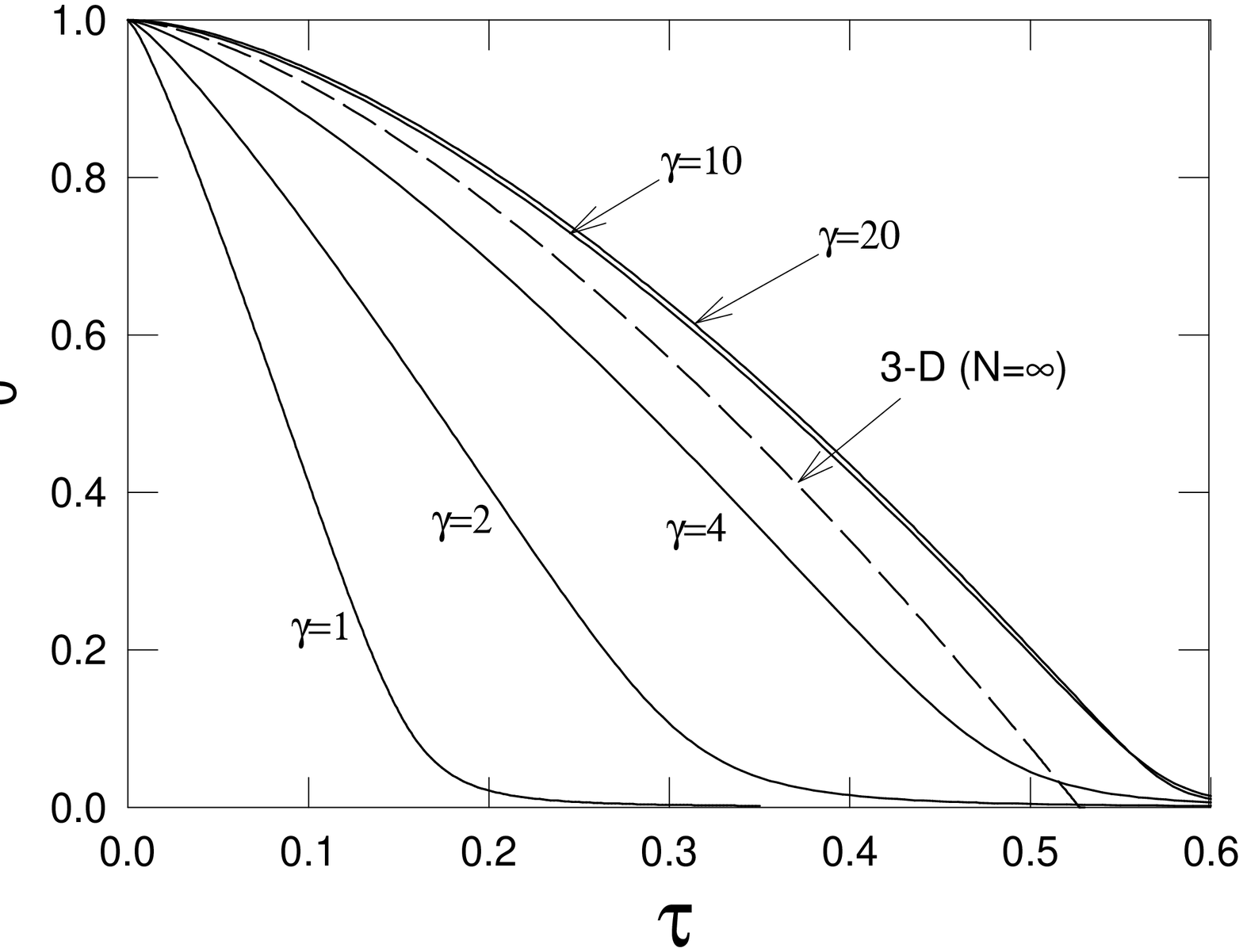} }    	
\end{center}
\vspace{2.3 in}
\label{figure3}
\vspace{6.0cm}
\centerline{}
\end{figure}
%.................................................................
%%%%%%%%%%%%% FIG. 4 %%%%%%%%%%%%%%%%%%%%%%%%%%%%%%%%%%%%%
\begin{figure}[p]
\begin{center}
\leavevmode
\hbox{	
\includegraphics{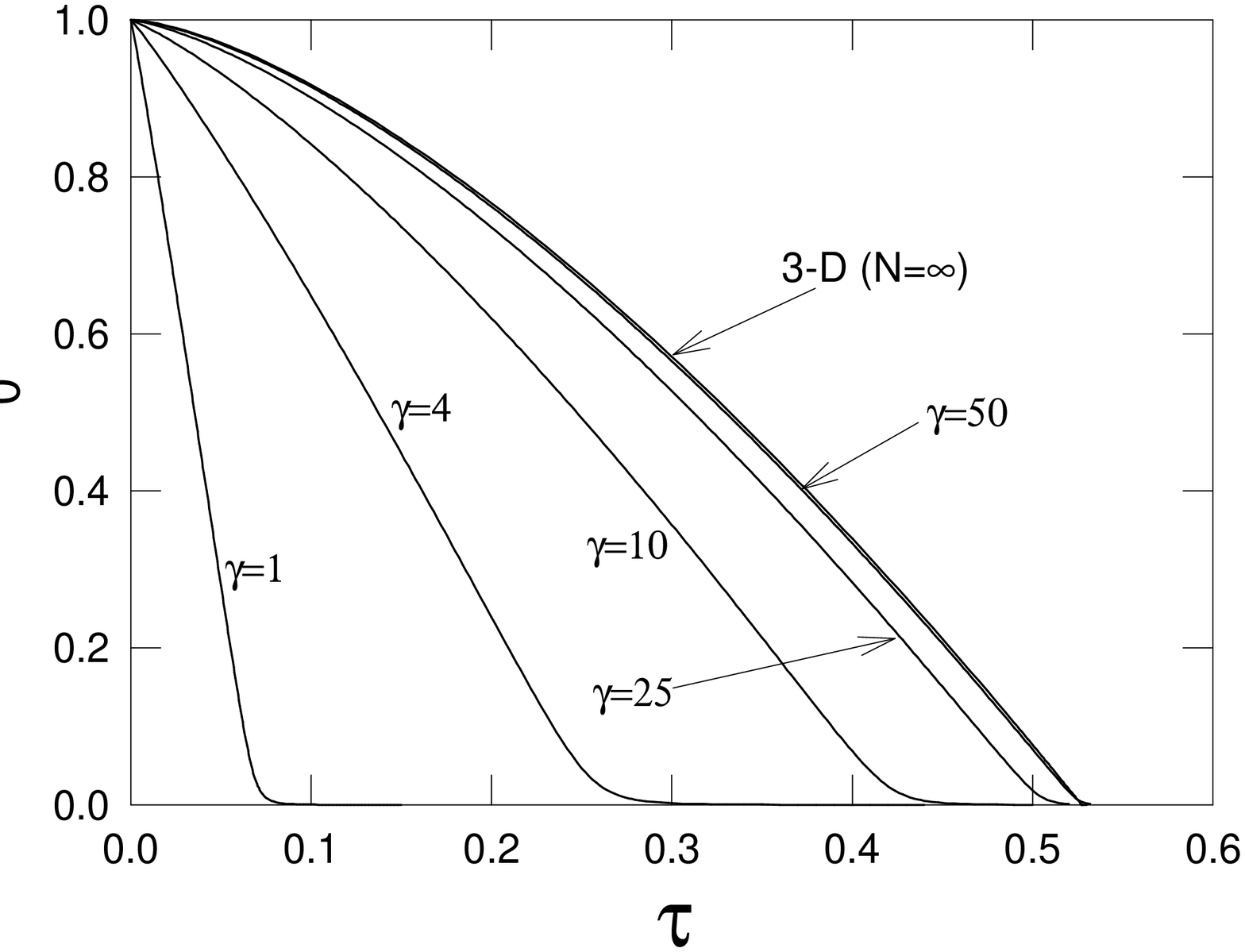} }    	
\end{center}
\vspace{2.3 in}
\label{figure4}
\vspace{6.0cm}
\centerline{}
\end{figure}
%%%%%%%%%%%%%%%%%%%%%%%%%%%%%%%%%%%%%%%%%%%%%%%%%%%%%%%%%%%%%%%%%%%%%%

\end{sloppypar}
\end{document}